# Interplay between superconductivity and ferromagnetism in epitaxial Nb(110)/Au(111)/Fe(110) trilayers

Hiroki Yamazaki,[1,*] Nic Shannon,[2,3] and Hidenori Takagi[1,2,3]

[1]*Magnetic Materials Laboratory, Discovery Research Institute,*
*RIKEN (The Institute of Physical and Chemical Research),*
*Wako, Saitama 351-0198, Japan*
[2]*Department of Advanced Materials Science,*
*Graduate School of Frontier Sciences, University of Tokyo, 5-1-5,*
*Kashiwanoha, Kashiwa, Chiba 277-8851, Japan*
[3]*CREST, Japan Science and Technology Agency, Kawaguchi 332-0012, Japan*

In order to clarify the influence of ferromagnetism on superconductivity through a normal-metal layer, the superconducting properties of epitaxial Nb(110)/Au(111)/Fe(110) trilayers were studied as a function of the thickness $t_{Au}$ of the intervening Au layer. Structural characterization of the samples revealed sharp interfaces, almost free from roughness. A strong suppression of the superconducting transition temperature $T_c$ was observed for $t_{Au}<10$ Å, implying a strong spin-polarization of the Au layer in the vicinity of the Au/Fe interface. A periodic change of $T_c$ with a period of ~21 Å (~9 atomic monolayer of Au) was observed for 20 Å$<t_{Au}<$104 Å. Neither the Fermi surface nesting of the normal metal layer, nor Fulde-Ferrel-Larkin-Ovchinnikov oscillations induced by a superconducting proximity effect in the ferromagnet can by themselves account for the observed period. These results suggest that a new and more subtle form of quantum interference occurs in very clean trilayer systems.

# I. INTRODUCTION

The properties of S/F (superconductor/ferromagnet) junctions and multilayers have been intensively studied from a standpoint of the interplay between superconductivity and ferromagnetism. Though in a special case, superconductivity and ferromagnetism can coexist and exhibit novel phenomena such as the π state in S/F/S junctions;[1-3] in general, a ferromagnet attached to a superconductor is expected to suppress the order parameter in the superconductor. For the experimental study of the interplay between superconductivity and ferromagnetism, the change of the superconducting properties as functions of the thicknesses of ferromagnetic and superconducting layer has been investigated for many S/F combinations: Nb/Fe,[4-6] Nb/Gd,[7,8] V/Fe,[9-11] V/V$_{1-x}$Fe$_x$,[12] Pb/Fe,[13] Al/Co,[14] and YBa$_2$Cu$_3$O$_{7-\delta}$/La$_{0.7}$Ca$_{0.3}$MnO$_3$.[15] In these multilayers, contrary to expectations, interface effects such as interface transparency and alloying at interfaces play considerable roles and veil the mechanism of the proximity effect in S/F heterostructures. The change of the magnetic properties of the ferromagnetic layer when its thickness is reduced has also become an obstacle for the interpretation of experimental results.

In the present study, to evade the difficulties mentioned above, we fabricated a series of S/N/F trilayers (N: normal metal), in which interface effects and the magnetic properties of the ferromagnetic layer are kept constant while only the thickness of N is changed. There is no direct contact between F and S, so that the interaction between F and S is expected to occur via the modification of the conduction electrons in N by means of the spin polarization and/or the proximity-induced superconductivity. Earlier studies of the Fe/Pt/Nb multilayers[16] confirm that $T_c$ can be strongly suppressed by the presence of a ferromagnetic layer. The present study goes further in using structurally well-defined samples to systematically investigate $T_c$ as a function of the thickness of N.

In a clean trilayer system, where the thickness of N is small compared with the superconducting coherence length and the mean free path for conduction electrons, proximity induced superconductivity and ferromagnetism couple directly. In this case, the superconducting transition temperature $T_c$ will depend very sensitively on the properties of the entire trilayer system, and not just on those of the superconducting and ferromagnetic layers. This raises the possibility of new and entirely quantum effects. We anticipate, for example, that there will be an RKKY-like oscillation in the effective magnetic exchange-interaction between F and S, analogous to that observed in some F/N/F trilayers and F/N multilayers.[17-19] Besides this, the superconducting proximity effect in S/F multilayers is known to exhibit Fulde-Ferrel-Larkin-Ovchinnikov (FFLO) oscillations. How these effects combine in a S/N/F trilayer is an open question.

## II. SAMPLE PREPARATION AND EXPERIMENTS

For the study of S/N/F trilayers, a series of Nb(110)/Au(111)/Fe(110) trilayers was prepared using a molecular-beam-epitaxy (MBE) machine (Eiko Co., Japan). In a growth chamber with a base pressure less than $10^{-10}$ Torr, the metals of Nb, Au, and Fe were evaporated in sequence by an electron beam to form trilayers on a substrate. For *in-situ* rate and thickness control, a calibrated quartz-crystal oscillator (INFICON) was used. The calibration factors had been reappraised taking the results of *ex-situ* reflection X-ray diffraction study into account. A single crystal of $Al_2O_3(11\bar{2}0)$ ($10 \times 10 mm^2 \times 0.3 mm$) was used as a substrate. This substrate was heat-cleaned at 750°C for 10 min in the growth chamber before beginning growth of the layers. First, keeping the substrate temperature at 750°C, a Nb layer of 288 Å was grown on the substrate with a rate of deposit 0.6±0.2 Å/sec. The thickness of the Nb layer was chosen to be of the same order as the superconducting coherence length $\xi(0) \sim 410$ Å (Ref. 20) of bulk Nb for clean-limit at absolute zero. Subsequently, the substrate was cooled down to 100°C then a Au layer ($t_{Au}$=0-104 Å) and an Fe layer (126 Å) were deposited with a rate of deposit 0.10±0.05 Å/sec. The trilayer was finally capped with a Au layer of 44 Å in order to avoid oxidization. Since the thickness of the Fe layer is effectively larger than the penetration depth of Cooper pairs into an Fe layer ($\xi_M \sim 12$ Å),[5] the Au capping layer does not affect superconductivity of the trilayer. A schematic diagram of the vertical section of the sample is shown in the inset of Fig. 1. A series of Nb(110)/Au(111) bilayer ($t_{Au}$=0-87 Å) was also prepared on the $Al_2O_3(11\bar{2}0)$ substrate as reference samples under the same preparation conditions as those of the trilayers.

Patterns of reflection high-energy electron diffraction (RHEED) were observed for the surface of each layer during the sample growth. Structural characterization was performed mainly by *ex-situ* small-angle and middle-angle reflection X-ray diffraction measurements using the X-ray diffractometer systems of MAC Science with Cu K$\alpha$ radiation and Philips MRD with Cu K$\alpha_1$ radiation. In order to study superconducting properties of the trilayers and the bilayers, magnetic susceptibility measurements were performed using a superconducting quantum interference device magnetometer (Quantum Design MPMS2). For some of the samples, resistivity measurements were also carried out to check the quality of the samples and to confirm consistency of $T_c$'s determined magnetically and resistively. A standard four-terminal configuration by a low frequency ac technique was used with a current density less than 1 A/cm$^2$. The current and voltage leads were attached to the sample surface by silver epoxy. Susceptibility measurements are in principle a better probe of bulk superconductivity than resistivity measurements, since the latter probe only the superconducting path with the highest $T_c$. Empirically, however, we found that the values $T_c$ obtained for the trilayers by the two different methods agreed to within experimental accuracy, confirming the homogeneity of the samples.

## III. STRUCTURAL CHARACTERIZATION

In the determination of layer structure, X-ray diffraction is a powerful technique. A typical pattern of the middle-angle 2θ-θ scan is shown in Fig. 1 for $t_{Au}$=26 Å. The sharp peak at ~38° is of substrate origin. For middle-angle scans, we see the presence of well resolved Laue oscillations from Nb as well as Fe, indicating that the structural coherence of the layers exists through the intervening Au layer. All the samples show the (110) growth of Nb and Fe with a structural coherence length, which was estimated from the peak width, corresponding to each layer thickness. The average difference of the lattice spacing $\Delta d=(d-d_{bulk})/d_{bulk}$, where $d_{bulk}$ is the lattice spacing of bulk material, is –0.97% and +0.17% for Nb(110) and Fe(110), respectively. Because of the close peak-position of Nb(110) and Au(111) and of the broadening of the Au(111) peak due to the small thickness of the Au layer, the Au(111) peak is not well resolved at middle angles. At large angles of 2θ~82°, however, the Au(222) and Nb(220) peaks are comparatively resolved for larger $t_{Au}$s. The rocking curves of the Nb(110) and Fe(110) Bragg peaks have typically a full-width-at-half-maximum (FWHM) of 0.05° and 0.9°, respectively, suggesting that the distribution of the (110) growth direction is small but is somewhat wider for Fe than for Nb.

In order to check the roughness of interfaces and each layer thickness, diffraction intensity measured in the small-angle 2θ-θ scans was fitted to the optical calculations using the profile-fitting program of SUPREX developed by Fullerton et al.[21] In Fig. 2, typical diffraction data for $t_{Au}$=17 Å (circles) and the result of fitting (solid curve) are shown. The calculation result is multiplied by 1/10 for clarity of comparison. Using the interface roughness $\sigma_n$ ($n$=0-4) and the layer thicknesses indicated in the inset, the experimental data are well reproduced by the calculation curve. Here the distribution of $d\rho(z)/dz$ in the vicinity of the interface is regarded as a Gaussian profile with a width of $\sigma_n$, where $\rho(z)$ is the density as a function of the position $z$ and the direction of $z$ is taken perpendicular to the interface. Judging from the small values of $\sigma_2$=1.9 Å (less than one-atomic-monolayer roughness) at Au/Fe and $\sigma_3$=0.0 Å at Nb/Au, the trilayer has considerably sharp interfaces at the bottom and top of the Au. High quality interfaces should provide the ideal stage for the study of the proximity effect in the S/N/F junction.

For the surface of the Au layer, a fine streak pattern of RHEED (Figs. 3(b) and 3(c)) was observed for all the Au-layer thicknesses of $t_{Au}$=4-104 Å. This indicates that epitaxial layer-by-layer growth occurs from the early stages of the Au-layer growth. The pattern of the Fe layer, on the other hand, exhibited dependence on $t_{Au}$. For $t_{Au}$<17 Å and $t_{Au}$≥17 Å, the patterns of Figs. 3(d) and 3(e) were observed, respectively. This change in the RHEED pattern of the Fe layer occurs suddenly at $t_{Au}$=17 Å implying the presence of a structural critical thickness of the Au layer when covered with the Fe layer. Structural change at $t_{Au}$=17 Å, however, is not detected by X-ray diffraction measurements. The diffraction pattern of Fig. 3(e) can be understood in terms of the twinning of the Fe-layer crystal around its <110> axis that is perpendicular to the sample plane. Because of this twinning, the average lattice strain in the Au layer, which is partly due to the lattice mismatch between Au and Fe, is likely somewhat different between $t_{Au}$<17 Å and $t_{Au}$≥17 Å.

## IV. SUPERCONDUCTING PROPERTIES

The conducting and superconducting properties of the Nb layer that is common to all the samples were investigated. First, the temperature dependence of the resistivity ρ was measured for a single Nb layer of 288 Å without any capping layer on it. It exhibits $T_c$=9.04±0.01 K (at 10% of the normal-state resistivity) with a transition width of 0.04±0.01 K (10-90% criterion) and the residual resistivity ratio RRR=ρ(300 K)/ρ(9.15 K)~10. The obtained $T_c$ agrees with $T_c^{on}$ to within experimental accuracy, where $T_c^{on}$ is the superconducting transition temperature defined as the onset point of the diamagnetic transition. A small reduction of ~0.2 K in $T_c$ from the bulk value (9.23 K) may be understood in terms of the reduced dimension of film. The electron mean free path $l^{Nb}$ is calculated from the product $<\rho_n^{Nb} l^{Nb}>$=3.75×10$^{-16}$ Ωm$^2$ for bulk Nb,[20] where $\rho_n^{Nb}$ is the residual resistance and the brackets mean averaging over the Fermi surface. The effective mean free path of $l_{eff}^{Nb}$~250 Å is obtained from $\rho_n^{Nb}$=1.5×10$^{-8}$ Ωm at 9.15 K (>$T_c$) using the substitution $<\rho_n^{Nb} l^{Nb}>$=$\rho_n^{Nb} l_{eff}^{Nb}$. Secondly, the superconducting coherence length in the Nb layer was determined from the measurements of the upper critical field $H_{c2\perp}(T)$ for the Nb[288 Å]/Au[$t_{Au}$] bilayers. Magnetic field was applied perpendicular to the sample plane. A linear temperature dependence: $H_{c2\perp}(T)=H_{c2\perp}(0)(1-T/T_c)$, was obtained near $T_c$ with $H_{c2\perp}(0)$=3.3, 3.1, 3.4, and 3.3 kOe for $t_{Au}$=10.4, 32.6, 60.9, and 87.0 Å, respectively. In the frame work of the Ginzburg-Landau theory, $H_{c2\perp}(0)$ is given by $H_{c2\perp}(0)=\phi_0/2\pi\xi_{GL}(0)^2$, where $\phi_0$ is the flux quantum and $\xi_{GL}$ is the Ginzburg-Landau coherence length. For the $H_{c2\perp}(0)$ values, a coherence length of $\xi_{GL}(0) \sim 320$ Å was obtained irrespective of $t_{Au}$. It is to be noted that $\xi_{GL}$ and $l_{eff}^{Nb}$ are in the same order of magnitude, indicative of an intermediate (between dirty and clean) regime of superconductivity, and are comparable to the thickness of the Nb layer ($t_{Nb}$=288 Å).

The residual resistivity of the Nb/Au bilayers was measured at 9.1 K ($>T_c$). The typical resistivity of the Au layer was estimated to be $\rho_n^{Au}$=3.3×10$^{-9}$ Ωm using a parallel resistor model, where the two layers of Nb and Au were treated as two parallel resistors, and a value of $\rho_n^{Nb}$=1.5×10$^{-8}$ Ωm, measured for a single Nb layer in the absence of Au (see above), was used for the Nb "resistor". The measured value of $\rho_n^{Au}$ corresponds to a very long mean free path of $l^{Au}$=92 Å×$[(r_s/a_0)^2/\rho_n^{Au}] \sim 2500$ Å, where $\rho_n^{Au}$ is measured in μΩcm and $r_s/a_0$=3.01 (ref. 22). We note that this is greater than the thickness of the Au spacer layer in our trilayer samples, and that charge transport through the spacer layer is therefore essentially ballistic.

For superconductivity and ferromagnetism to couple directly, it is necessary that there be a finite superfluid density throughout the Au spacer, so that the cooper pairs of Nb can "feel" the exchange field of Fe. For this we require: $t_{Au} \leq \xi_N$ and $t_{Nb} \leq \xi_{GL}$, where $\xi_N$ is the characteristic coherence length of the normal state. In the strict sense, these conditions are loose for the so-called "Cooper limit" ($t_{Au} << \xi_N$ and $t_{Nb} << \xi_{GL}$), but they are sufficient from the experimental point of view. In order to confirm that these conditions are realized in our trilayer sample, we studied the dependence of $T_c$ on $t_{Au}$ for the Nb/Au bilayers in advance.

The Cooper limit holds if that the electrons probe the entire bilayer and experience a pairing interaction which is an average between the two materials. For a strong-coupling superconductor like Nb, $T_c$ is given by the McMillan formula:[23]

$$T_c = \frac{\omega_D}{1.45} \exp\left[-\frac{1.04(1+\lambda_{eff})}{\lambda_{eff} - \mu^*(1+0.62\lambda_{eff})}\right], \quad (1)$$

where $\omega_D$ is the Debye temperature and $\mu^*$ is the Coulomb pseudopotential, and the electron-phonon coupling constant $\lambda$ has been replaced by an average quantity of $\lambda_{eff}$. Following de Gennes,[24] we write: $\lambda_{eff}=(\lambda_{Nb}N_{Nb}t_{Nb}+\lambda_{Au}N_{Au}t_{Au})/(N_{Nb}t_{Nb}+N_{Au}t_{Au})$, where $N$ is the density of states at Fermi level, $t$ is the layer thickness, and the subscripts denote the corresponding materials. This model assumes that the electrons spend $N_{Au}t_{Au}/(N_{Nb}t_{Nb}+N_{Au}t_{Au})$ of their time in the normal metal of Au and $N_{Nb}t_{Nb}/(N_{Nb}t_{Nb}+N_{Au}t_{Au})$ in the superconducting Nb.

The result of the fitting of the Eq. (1) to the experimental data is shown in Fig. 4, where the on-set transition temperature $T_c^{on}$ determined magnetically is shown as a function of $t_{Au}$ for the Nb[288 Å]/Au[$t_{Au}$] bilayers. In the fitting the data, we have used the single fitting parameter $\lambda_{Au}$, together with a value $\lambda_{Nb}$=0.76 determined from $T_c^{on}$=9.04 K at $t_{Au}$=0 Å. This value of 0.76 for $\lambda_{Nb}$ is in good agreement with $\lambda_{Nb}$=0.82 of bulk Nb.[23] The known values of Nb and Au were used for other quantities: $\omega_D$=277 K (bulk Nb),[23] $\mu^*$=0.11 (bulk Nb),[25] $N_{Nb} \propto \gamma_{Nb}$=7.186×10$^3$ erg/cm$^3$K$^2$ (bulk Nb),[26] $N_{Au} \propto \gamma_{Au}$=0.715×10$^3$ erg/cm$^3$K$^2$ (bulk Au),[26] and $t_{Nb}$=288 Å, where $\gamma$ is the electron heat capacity constant. The best fit was obtained for $\lambda_{Au}$=0.17±0.04. This compares reasonably with values $\lambda_{Cu}$=0.25 and $\lambda_{Ag}$=0.2 previously obtained from other noble metal bilayer systems.[27,28] We anticipate that $\lambda_{Cu} > \lambda_{Ag} > \lambda_{Au}$, since $\lambda$ is a monotonically increasing function of $T_c$, which in turn decreases with the atomic mass. We see that, using the above reasonable parameters, the McMillan expression within the framework of the de Gennes model reproduces the experimental data well. This analysis shows that the superconducting order parameter extends throughout the Au layer, i.e. that $t_{Au} \leq \xi_N$ holds up to $t_{Au}$~100 Å suggesting $\xi_N$ >100 Å. By extension, the same must be true for the corresponding Nb/Au/Fe trilayers. Since the crucial Au/Fe interface is confirmed to be almost free of roughness in the trilayers, all loss of coherence and/or the breaking of the Cooper pairs in the Au layer can be uniquely attributed to the exchange field of the Fe layer.

Typical temperature dependence of the normalized magnetic susceptibility $\chi/|\chi_{(5K)}|$ for the trilayers of $t_{Au}$=82.7-104.4 Å is shown in Fig. 5, where $\chi_{(5K)}$ is the susceptibility at 5 K. The measurements were carried out for the warming (after zero-field cooling in $|H|$<0.002 Oe) and cooling procedures in a magnetic field of 2 Oe applied perpendicular to the film plane. A clear indication of superconductivity can be seen by the diamagnetic response. We note in passing that the magnitude of the remnant magnetization of the Fe layer did not affect the temperature dependence of $\chi$. This was confirmed by a comparison between the two methods of producing zero-magnetic field at the sample position: (i) using an oscillation mode (using oscillating current of the superconducting magnet) and (ii) using a monotonic reduction of field to zero after applying a field of 1 T. The method of (ii) produced much more remnant magnetization in the Fe layer than (i), but the $\chi$-$T$ curves obtained for the two methods corresponded within experimental accuracy. In both cases, the direction of the remnant magnetic moment was in the film plane.

The dependence of $T_c$s ($T_c^{on}$ and $T_c^{50\%}$) on $t_{Au}$ is shown in Fig. 6 together with $T_c^{on}$ of the Nb/Au bilayers, where $T_c^{50\%}$ is the temperature at which $\chi$ has a 50% value of $\chi_{(5K)}$.[29] Relatively broad widths of the transition are obtained in susceptibility measurements - $\Delta T_c \propto (T_c^{on} - T_c^{50\%})$~1.2 K and ~0.6 K for $t_{Au}$<17 Å and $t_{Au} \geq$17 Å, respectively, as compared to $(T_c^{on} - T_c^{50\%})$~0.2 K in a single Nb layer. The jump in $T_c$ and $\Delta T_c$ at $t_{Au}$=17 Å (~7 atomic monolayer (ML)) corresponds to the change in the RHEED pattern of the Fe surface. As mentioned in the previous section, the lattice strain in the Au layer for $t_{Au}$<17 Å is likely somewhat different from that for $t_{Au} \geq 17$ Å. From a quantitative standpoint, the behavior of $T_c(t_{Au})$ should be treated separately for $t_{Au} < 17$ Å and $t_{Au} \geq 17$ Å.

Comparison of $T_c^{on}(t_{Au})$ between the bilayers and the trilayers immediately tells us that superconductivity is obviously suppressed in the trilayers. Considerable reduction of $T_c$ is observed particularly for $t_{Au} < 10$ Å in the trilayers. At $t_{Au} = 0$ Å, when direct contact between Nb and Fe occurs, the reduction of $T_c^{on}$ by 3.7 K is obtained very likely due to the proximity to the Fe layer. We understand that the strong spin-polarization of the conduction electrons due to the Fe layer extends ~10 Å inside the Au layer, so that, when $t_{Au} < 10$ Å, the Nb layer encounters the strong polarization, resulting in the reduction of $T_c$ by the Cooper-pair breaking. This explanation is quite consistent with the result of the Mössbauer spectroscopy for the Au/M (M=Fe, Co, and Ni) multilayers using $^{119}$Sn probes: the penetration depth of the spin polarization of conduction electrons ($\xi_{sp}$) extends over ~10 Å into the Au layer from the Au/M interface.[30]

The $T_c$ of the bilayers monotonically decreases with $t_{Au}$ due to the superconducting proximity effect, while the trilayers, as a whole, show completely opposite behavior. The lowest $T_c$ is observed for $t_{Au} = 0$ Å, and $T_c$ increases as $t_{Au}$ increases. Though the $T_c$ of the trilayers seems to be saturated for the $t_{Au}$s more than about 20 Å, the weak interplay between the Fe layer and the Nb layer remains even for $t_{Au} > 20$ Å because the $T_c$s of the trilayers are still lower than those of the bilayers. The exchange field of the Fe layer will clearly affect the Cooper pairs in the neighbouring Au layer. In the light of the Cooper criterion, above, this effect must propagate to the Nb layer.

In the trilayers, as hitherto mentioned, the conduction electrons in the Au layer are under the influence of not only the proximity-induced superconductivity from the Nb layer but also the spin polarization of the Fe layer. The characteristic length scale of the interplay between the Nb and the Fe layer via the Au layer, however, is not limited to $\xi_{sp}$. As we can see in Fig. 6, the $T_c^{on}$-$t_{Au}$ curves of the bilayers and the trilayers seem to converge at some value of $t_{Au}$ larger than 100 Å, where the effect of the Fe layer on the Nb layer should be completely "screened" by the intervening Au layer. It is quite likely that $\xi_N$, which is estimated to be more than 100 Å, restricts the length scale of the interplay between Nb and Fe.

In the weakly coupled region for $t_{Au} > 20$ Å, unexpectedly, we clearly see a periodic change of $T_c^{on}$ and $T_c^{50\%}$ with a period of ~21 Å (~9 ML of Au). The vertical broken lines in the figure are drawn at intervals of 21 Å. The period and amplitude of the oscillation (about 5% of $T_c$) remain clearly defined, irrespective of whether $T_c^{on}$ or $T_c^{50\%}$ is used. Indeed, this periodic change can be unambiguously identified in the raw $\chi$-$T$ data shown in Fig. 5. As $t_{Au}$ increases from 82.7 to 100.1 Å, the curve shifts to the higher temperature; as $t_{Au}$ increases from 100.1 to 104.4 Å, it shifts to the lower temperature, however. It should be emphasized that the shift occurs parallel to the temperature-axis. We also confirmed that the period and phase of $T_c(t_{Au})$ do not exhibit field-dependence for $H \leq 100$ Oe; although the curve of $T_c(t_{Au})$ shifts downward as $H$ increases, the peaks and dips hold their positions in $t_{Au}$. Since the periodic oscillation in $T_c(t_{Au})$ is not accompanied by any change in crystal structure, we conclude that it is of intrinsic, electronic origin. It may be interesting to infer that the minimum of the oscillating $T_c$ is always located at $n \times 21$ Å ($n$: integers) suggesting that there is a hidden first minimum at $t_{Au}=0$ Å, though the presence of the $T_c$ jump at $t_{Au}=17$ Å makes it ambiguous.

## V. DISCUSSION

The most striking feature of our results is the presence of a well-defined periodic oscillation in its superconducting transition temperature of a clean, epitaxialy grown Nb/Au/Fe-trilayer system as a function of the thickness of the normal metal layer. The simplest possible explanation for such an oscillation in $T_c$ would be a periodic Friedel-type oscillation in the density of electrons in the Au. Such an oscillation could be triggered by the abrupt boundary between Nb and Au, and would have a period tied to the nesting vectors of the Fermi surface of Au (discussed below). Since the density of states of the Au would be modulated by such a Friedel oscillation, the superconducting transition temperature of the Nb/Au/Fe trilayer would vary with the same period. However we can almost certainly rule out this possibility on the basis of our results for Nb/Au bilayers. In these, $T_c$ exhibits a simple monotonic decay, which can be understood in terms of the superconducting proximity effect of the Nb. Quantitative fits to $T_c$ can be obtained assuming a constant density of states on the condition that the superconducting order parameter extends throughout the neighboring Au layer, and that the density of state in Au is constant. Clearly, the oscillations in the $T_c$ of the trilayer system must have something to do with the ferromagnetism of the additional Fe layer.

The simplest explanation which takes this magnetism into account is to assume that an oscillating RKKY-like coupling exists between the magnetic Fe and superconducting Nb layers, mediated by the conduction electrons of the Au layer. This effect is known to occur in F/N multilayer systems, where the ferromagnetic layer induces a spin density wave in the normal metal, with a characteristic period determined by the relevant nesting vector of its Fermi surface.[31] Direct *ab initio* calculations of the Fermi surfaces of noble metal spacers give rise to nesting vectors with wavelengths of order a few monolayers (ML), irrespective of the precise details of the spacer layer;[32] the corresponding oscillation in $T_c$ would have only half of this period since the pair breaking effect in the Nb is independent of the sign of the exchange field. Clearly this period is too short to account for the oscillations seen in our trilayer sample which have a period of ~9 ML, i.e. 21 Å. In fact, as was realized by Bruno and Chappert, umklapp processes must be taken into account when determining the true period of oscillations in a finite size system with a lattice.[31] However, even allowing for these, the predicted oscillation period of 4.83 ML = 11.35 Å is still much too short to explain our data.

The third explanation we consider builds on established results for S/F multilayers. In these, the proximity effect of the superconductor in the neighboring ferromagnetic layer is strongly suppressed by the exchange field. Moreover, the rapidly decaying superconducting order parameter in the ferromagnet acquires a periodic oscillation originally predicted by Fulde, Ferrel, Larkin and Ovchinnikov in the context of bulk ferromagnetic superconductors.[33,34] The period of this oscillation is given by $1/Q = v_F/2h_E$, where $v_F$ is the Fermi velocity and $h_E$ the exchange field of the ferromagnetic layer.[35] In a clean trilayer system, such FFLO oscillations in the superfluid density of the ferromagnetic layer would enter the boundary conditions for superconductivity in the normal layer, and so influence $T_c$. However, once again, this mechanism fails to offer a quantitative explanation of our results. Taking values of $v_F=1.98\times10^6$ m/s (ref. 22) and $h_E$~2 eV (ref. 36) from photoemission and band structure studies of bulk iron, we find a period of ~6 Å, substantially shorter than the ~21 Å period of the $T_c$ oscillations in our trilayer system.

We note that it would be possible to explain the long period of these oscillations seen in experiment in an *ad hoc* way by assuming an anomalously low value of the exchange splitting $E_{ex}\approx100$ meV. Extremely low values of exchange splitting have previously been inferred for layered structures.[37] One possible mechanism for this in a S/N/F trilayer might be a weaker, induced ferromagnetism in the metallic spacer. In the absence of a complete theoretical understanding of the trilayer system it is difficult to rule out such an effect categorically. However, since the long mean free path implies that charge transport through the spacer layer is ballistic in nature, we do not consider this to be a probable explanation.

Clearly more theoretical work is needed before any definitive conclusion can be drawn about the experiments described in this paper. However, on the basis of the evidence available, we conclude that neither the FFLO oscillations seen in S/F bilayers, nor the oscillating RKKY-type interactions in a F/N multilayer can by themselves explain our results.

We anticipate that both effects *will* be present in a trilayer system, and a theory is needed which takes both into account simultaneously.  In particular, the clean, epitaxial nature of our samples and the high quality of the interfaces between the different metal layers raises the possibility of new quantum interference effects, in which novel bound states form in the noble metal layer.  At present, full many-body calculations for trilayers are not sufficiently advanced to take all of these different pieces of physics into account.

Sample quality may well prove to be the limiting factor in looking for such quantum effects.  Our preliminary studies of Nb/Pt/Fe trilayers do *not* show periodic oscillations in the superconducting transition temperature.  However in these systems, because of the poor "wetting" of Pt on Nb, we have not been able to fabricate interfaces between the Nb and Pt layers of equal quality to those between the Nb and Au layers discussed above.

Finally some comments are due on the Nb/Au/CoFe trilayers recently studied by Kim *et al.*[38]  These authors also found a similar periodic change of $T_c$ as a function of the thickness of the Au layer.  For Au layer thicknesses between 200 and 1100 Å, their trilayers exhibit a ~216 Å period, one order of magnitude larger than that found in our Nb/Au/Fe samples.  Theoretical studies of magnetic multilayers suggest that disorder tends to act as a low pass filter, suppressing short period oscillations.[31, 39, 40]  However in the absence of a complete theoretical understanding of the trilayer system problem, it is not clear whether this dramatic difference in period is intrinsic, or originates in different sample-preparation techniques - MBE for the Nb/Au/Fe trilayers as opposed to sputtering for the Nb/Au/CoFe.

## VI. CONCLUSION

The interaction between ferromagnetism and superconductivity, mediated by the conduction electrons of a normal metal, was studied in an epitaxial Nb(110)/Au(111)/Fe(110) trilayers of high quality.  A strong suppression of superconductivity occurred for Au-layer thicknesses of less than ~10 Å (~4 ML), which we attribute to the spin-polarization of conduction electrons in the vicinity of the Au/Fe interface.  In addition, a marked oscillation in $T_c$ with a period of ~21 Å was observed for $t_{Au}$>20 Å.  This oscillation cannot be understood either in terms of Fermi surface effects in Au, or of FFLO oscillations in Fe.  Our results suggest a new role for quantum coherence in clean multilayer samples, and motivate a thorough reexamination of the theory of S/N/F trilayers.

## ACKNOWLEDGEMENTS

The authors thank Prof. N. Hosoito for his useful information on the Mössbauer

**FIGURE CAPTIONS**

FIG. 1.   Typical reflection X-ray diffraction pattern of middle-angle 2θ-θ scan for the Nb[288 Å]/Au[26 Å]/Fe[126 Å] trilayer with Cu K$\alpha_1$ radiation.   The Al$_2$O$_3$(11$\bar{2}$0) peak is from the substrate.   Inset: schematic diagram of a vertical section of the sample.

FIG. 2.   Typical reflection X-ray diffraction pattern of small-angle 2θ-θ scan for the $t_{Au}$=17 Å sample.   The solid curve corresponds to the optical calculation result fitted to the experimental data using the profile-fitting program of SUPREX developed by Fullerton *et al*.[21]   The calculation result is multiplied by 1/10 for clarity of comparison.   Inset indicates the parameters (layer thicknesses and interface roughness $\sigma_n$ ($n$=0-4)) used in the calculation.

FIG. 3.   Reversal images of typical RHEED patterns obtained in the growth process of the Nb[288 Å]/Au[$t_{Au}$]/Fe[126 Å] trilayers.   The direction of the incident electron beam is parallel to <0001> of the Al$_2$O$_3$(11$\bar{2}$0) substrate.   The surface of the Fe layer exhibits the pattern of (d) ($t_{Au}$<17 Å) or (e) ($t_{Au}$≥17 Å).

FIG. 4.   Superconducting transition temperature $T_c^{on}$ (on-set temperature) as a function of the Au-layer thickness $t_{Au}$ for the Nb[288 Å]/Au bilayers.   The solid curve corresponds to the McMillan expression fitted to the experimental data using the effective electron-phonon coupling $\lambda_{eff}$ within the framework of the de Gennes model.[24]

FIG. 5.   Typical temperature dependence of normalized magnetic susceptibility $\chi/|\chi_{(5K)}|$ for the Nb[288 Å]/Au[$t_{Au}$]/Fe[126 Å] trilayers with $t_{Au}$=82.7-104.4 Å, where $\chi_{(5K)}$ is the susceptibility at 5 K.   Measurements were carried out for warming (after zero-field cooling) and cooling procedures in a magnetic field of 2 Oe applied perpendicular to the film plane.

FIG. 6.   Superconducting transition temperature $T_c$ ($T_c^{on}$ and $T_c^{50\%}$) as a function of the Au-layer thickness $t_{Au}$ for the Nb[288 Å]/Au[$t_{Au}$]/Fe[126 Å] trilayers.   The error of $T_c$ is within each symbol.   The vertical broken lines are drawn at intervals of 21 Å (~9 ML of Au).   The data of the Nb[288 Å]/Au[$t_{Au}$] bilayers and its theoretical fit (solid curve) are also indicated for comparison.

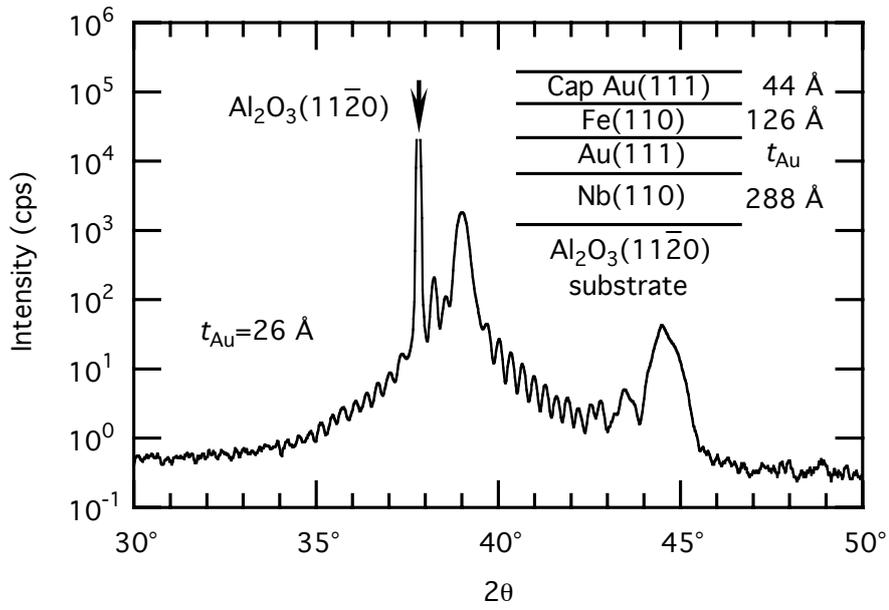

Fig. 1     H. Yamazaki *et al.*     P. R. B

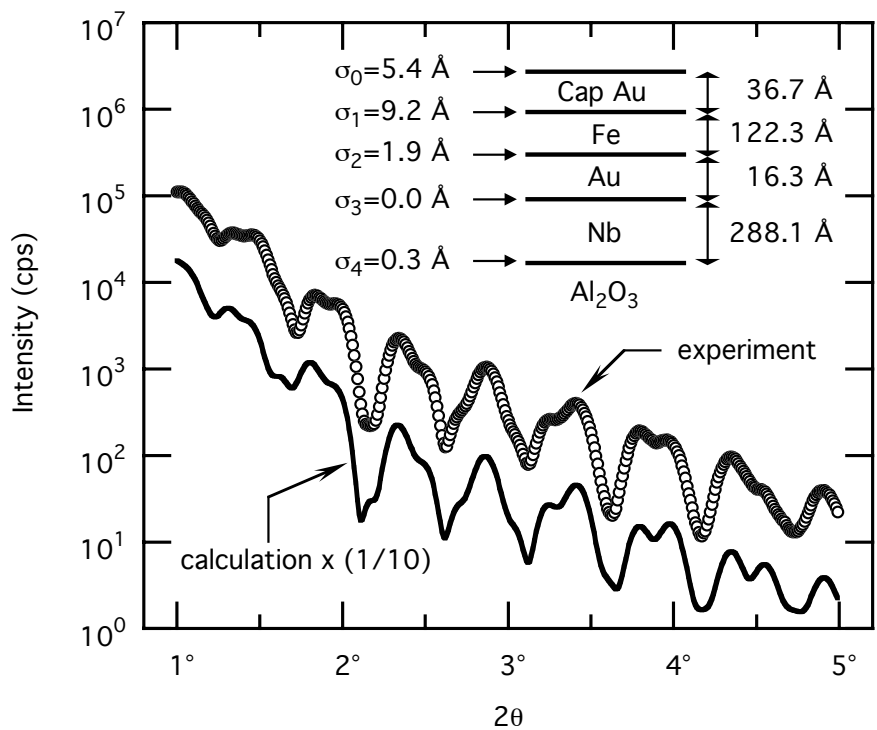

Fig. 2    H. Yamazaki *et al.*    P. R. B

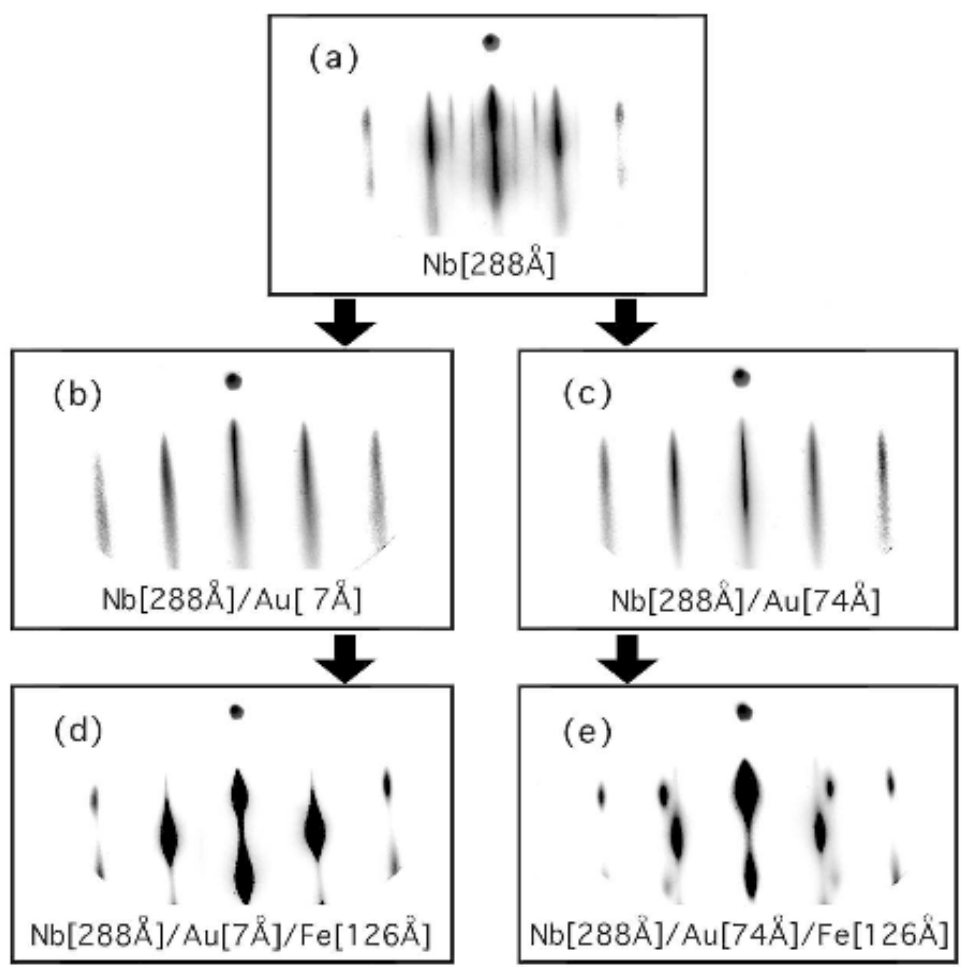

Fig. 3    H. Yamazaki et al.    P. R. B

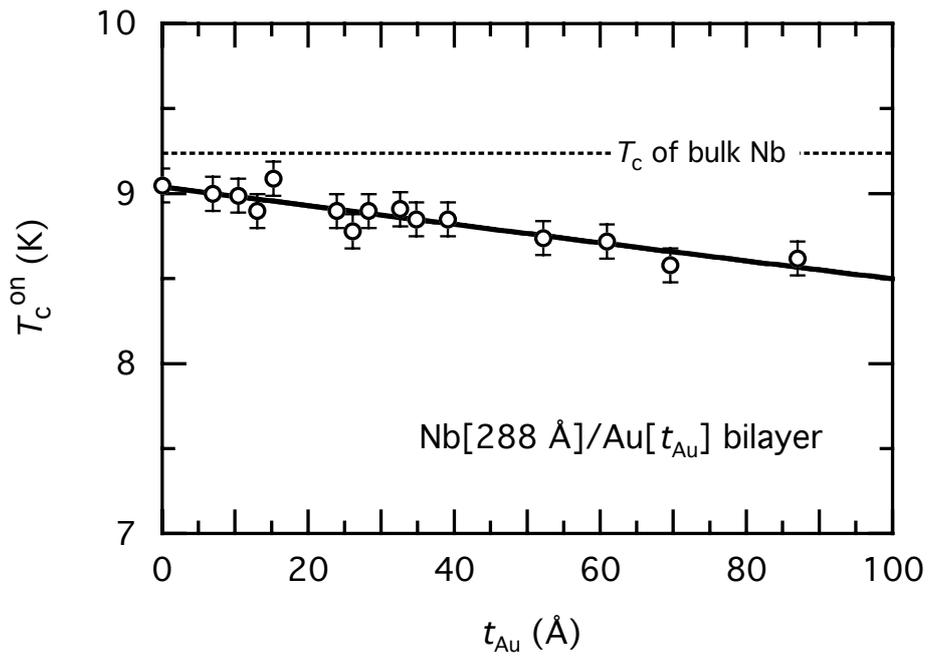

Fig. 4    H. Yamazaki *et al.*    P. R. B

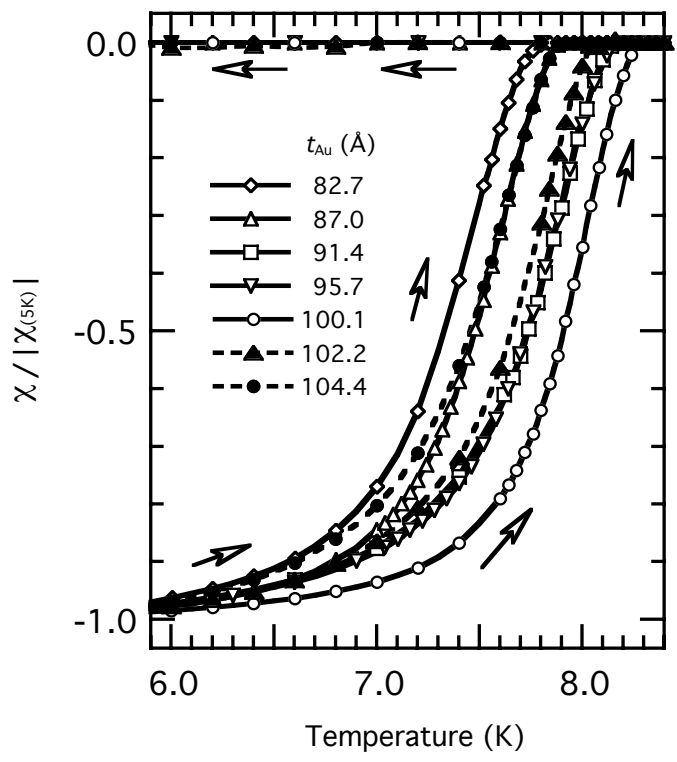

Fig. 5     H. Yamazaki *et al.*     P. R. B

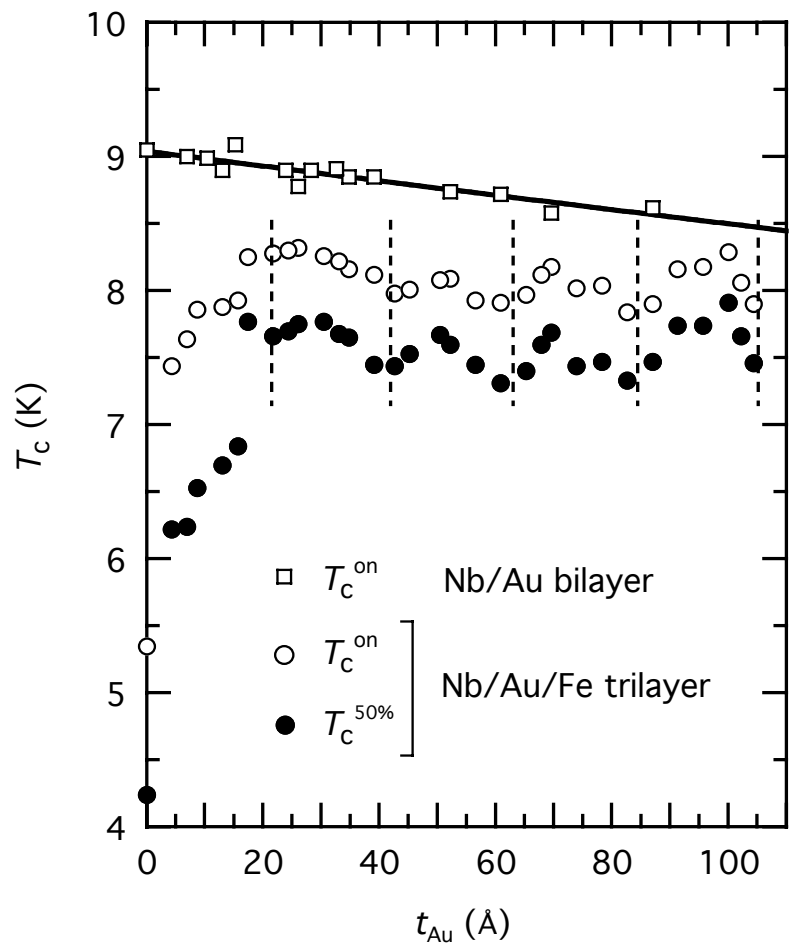

Fig. 6    H. Yamazaki *et al.*    P. R. B